\documentstyle[12pt]{article}

\renewcommand{\thefootnote}{\fnsymbol{footnote}}
\newcommand{\EQ}{\begin{equation}}
\newcommand{\EN}{\end{equation}}
\newcommand{\bea}{\begin{eqnarray}}
\newcommand{\ena}{\end{eqnarray}}
\newcommand{\vs}[1]{\vspace{#1 mm}}

\renewcommand{\a}{\alpha}
\renewcommand{\b}{\beta}

\newcommand{\uda}{\nearrow \kern-1em \searrow}

\newcommand{\w}{\wedge}
\newcommand{\cZ}{{\cal{Z}}}
\newcommand{\cG}{{\cal{G}}}
\newcommand{\cM}{{\cal{M}}}
\begin{document}

\topmargin 0pt
\oddsidemargin 5mm

\begin{titlepage}
\setcounter{page}{0}
\begin{flushright}
OU-HET 220 \\
July, 1995
\end{flushright}

\vs{10}
\begin{center}
{\Large Calabi-Yau Compactification of Type-IIB String  \\ and a Mass Formula
of the Extreme Black Holes}
\vs{10}

{\large
Hisao
Suzuki\footnote{e-mail address: suzuki@phys.wani.osaka-u.ac.jp}}\\
\vs{8}
{\em Department of Physics, \\
Osaka
University \\ Toyonaka, Osaka 560, Japan} \\
\end{center}
\vs{10}

\centerline{{\bf{Abstract}}}
Recently proposed mechanism  of the black hole condensation at conifold
singularity in type II string is an interesting idea from which we can
interpret the phase of the universal moduli space of the string vacua. It
might also be expected that the true physics is on the conifold singularity
after supersymmetry breaking. We derive a mass formula for the extreme black
holes caused by the self-dual 5-form field strength, which is stable and
supersymmetric. It is shown that the formula can be written by the moduli
parameters  of Calabi-Yau manifold and can be calculated explicitly.

\end{titlepage}
\newpage
\renewcommand{\thefootnote}{\arabic{footnote}}
\setcounter{footnote}{0}

There are huge number of consistent string theories in four dimensions.
Unification of the string vacua is required to recover the predictive power for
physics. Recently, an interesting observation was given by
Strominger\cite{Strominger} who argued that the conifold singularity may be the
key to understand the physics. Motivated by the argument given by Seiberg and
Witten\cite{SeibergWitten}, he interpreted the conifold singularity as a point
where black holes become massless and probably the physics choose the point
after supersymmetry breaking. The Wilsonian effective action acquires the
logarithmic term at conifold singularity and  induces such terms in the period
matrix which coincides with the classical calculation\cite{Strominger}. On the
other hand, it is known by mathematicians that we can glue together different
Calabi-Yau space at conifold singularity.  In Ref.\cite{GMS},  physical
understanding of such phase transition   through conifold singularity was
advocated. At present, the true physical process  has not fully understood  but
t!
 he argument is very interesting.

Motivated by these fascinating idea, we construct a mass formula of the
classical extreme black holes, which should be an fundamental formula for
Calabi-Yau compactification of the type II string. In type IIB string such
black holes are constructed by the self-dual 5 form field strength in ten
dimensions\cite{HS}. After compactification, the 5-form turns to the field
strength of the $U(1)$ Yang-Mills fields and the  black holes constructed by
these gauge field are stable and supersymmetric when other fields are neutral
with respect to these charges\cite{GH}. The masses of such black holes are
expected to be the function of the moduli\cite{Strominger}.
The mass formula will turn out to be different from the generalized BPS mass
formula for non-aberian dyon solutions\cite{CDFV}. More generally, the $d$-form
field strength in $2d$ dimensions contains the information of the internal
space  in the mass formula of the black holes. Therefore, it will be convenient
to treat the example in  dimensions lower that ten.

To begin with, we are going to consider torus compactification of the self-dual
3-form field strength in 6 dimensions. We denote the field strength by $H =
H_{\mu\nu\rho} dx^{\mu}\w dx^{\nu}\w dx^{\rho}$. By the $T^2$ compactification,
the internal space can be identified by
\bea
x = x + R_1, \qquad y = y + R_2,.
\ena
 and corresponding cycle can be called A-cycle and B-cycle respectively.
 We take the one-form basis $\alpha, \beta$ such that
\bea
\int_A \alpha = 1, \int_B \beta = -1, \int_{T^2}\alpha \w \beta =1.
\ena
The explicit form of the one forms are
\bea
\a = {dx \over R_1}, \b = -{dx \over R_2}.
\ena
The holomorphic $(1,0)$ form $\Omega = dx + idy$ can be written by these basis
as
\bea
\Omega = R_1 \a -i R^2 \b,
\ena
and satisfies
\bea
\int_A \Omega \equiv \cZ = R_1, \qquad \int_B \Omega \equiv \cG = -i R_2.
\ena
We expand the antisymmetric tensor fields as
\bea
H = F \w \a + G\w \a,
\ena
where $F$ and $G$ are the two form field strength with respect to space-time
and we have omitted the scalar component.
Then the action can be written as
\bea
\int_{T_2 \times M^4} H*H &=& \int F\w *F \a \w *\a + \int G\w*\w \b \w *\w
 \nonumber\\
&=& Vol(T^2)[{1 \over R_1^2} \int_{M^4} F\w * F + { 1 \over R_2^2} \int_{M^4} G
\w *G],
\ena
where $Vol(T^2) = R_1R_2$ and  this  volume factor of the internal space is
common to all terms of the action so that we can omit it.
The relation $(4)$  implies that the coupling constant depends explicitly on
the moduli parameter of the internal space. The limit $R_1 \rightarrow \infty$
can be regarded as the strong coupling limit.
As a classical solution, We require  the magnetic charge to be  quantized;
\bea
\int_{\a \times S^2} H = n C_g, \quad \int_{\b \times S^2}H = m C_g,
\ena
where the elemental charge $C_g$ can not be determined by our argument. The
quantum number $m$ and $n$ may be regarded as winding numbers with respect to
the internal space.
For the metric,
we take Reisner-Nordstrom solution. The condition $(8)$ implies that magnetic
component of the antisymmetric tensor can be written as
\bea
F_m = C_g n \sin\theta d \theta \w d \varphi, \quad G_m = - C_g m \sin\theta d
\theta \w d \varphi.
\ena
 Because of the self duality of $H$, the electric component can be obtained by
adding hodge dual of the tensor;
\bea
H = F_m \w \a + G_m \w \b + *F_m \w *\a + *G_m \w *\b.
\ena
In the case of the extreme black hole, the mass can be written by the sum of
the electric and magnetic charge as
\bea
M^2 = 2C_g^2({n^2 \over R_1^2} + {m^2 \over R_2^2}),
\ena
where the factor 2 comes from the fact that the contribution of the  electric
fields and the magnetic field are identical due to the electro-magnetic
duality.
When we denote the undetermined constant as $C_g = Vol(T^2) v = R_1 R_2
v/\sqrt{2}$, the mass of the extreme hole can be written as

\bea
M^2 = v^2 \vert \cZ m - \cG n \vert^2,
\ena
which is exactly the BPS formula\cite{SeibergWitten,CDFV} This formula  is
basically for dyon solutions in Non-Abelian theories.
You can easily generalize the construction to the torus compactification of the
5-form field strength in ten dimensions. But you will find that the formula
differs from that obtained in Ref.\cite{CDFV} by the special geometry.  This
fact implies that the mass formula is not unique, which may be caused by the
fact that the the elementary quantity is not $M$ but $M^2$ which can contain
tensor with respect to $Sp(n,Z)$. Before discussing Calabi-Yau compactification
of the 5-form field strength in ten dimensions,  we will derive some formula of
Calabi-Yau manifolds.

On the Calabi-Yau manifold $\cM$, we take a canonical homology basis for
$H_3(\cM;Z)$ as ${A^a,B_b}, a,b = 0,1,...,b_{2,1}$ and let $(\a_a,\b^b)$ be the
dual cohomology basis such that\cite{CGH}
\bea
\int_{A^a} \a_b = \int_\cM \a_a \w \b^b = \delta_a^b, \quad \int_{B_a} \b^b =
\int_\cM \b^b \w \a_a = - \delta_a^b.
\ena
Note that  holomorphic 3-form $\Omega$ is also the element of $H^3(\cM,Z)$. We
define the period of $\Omega$ as
\bea
z^a = \int_{A^a} \Omega, \qquad \cG_a = \int_{B_a} \Omega.
\ena
Therefore, we can express $\Omega$ as
\bea
\Omega = z^a \a_a - \cG_a(z)\b^a.
\ena
In Ref.\cite{BG}, it is shown that the complex structure of $\cM$ can be
completely determined by the choice of $z^a$, so that $\cG_a$ can be regarded
as a function of $z^a$. For the application to the algebraic geometry, it will
be convenient to
take $z^a$ as a function of other moduli parameters $t^a$\cite{BG}, but here
for simplicity, we take $z^a$ as elemental variables. Note that the holomorphic
3-form $\Omega$ is defined up to constant, the same must be required for $z^a$
and $\cG_a$;
\bea
\cG_a(\lambda z) = \lambda \cG_a(z),
\ena
from which we have
\bea
\cG_a = z^b\partial_b\cG_a.
\ena

By Kodaira's theorem\cite{BG}, the infinitesimal variation of the holomorphic
3-form with respect to the moduli parameters can be expanded by the
$(3,0)-$form and $(2,1)-$forms. Therefore,
\bea
\int_\cM \Omega \w \partial_a\Omega =0,
\ena
which implies $2\cG_a = \partial_a(z^b\cG_b)$. Therefore, $\cG$ can be written
as
\bea
\cG_a = \partial_a \cG.
\ena
The vector space $H^{(3,0)} \oplus H^{(2,1)}$ can be spaned by
\bea
\omega_a = \partial_a \Omega = \alpha_a - \partial_a \partial_b \cG \beta^b.
\ena
We hereafter denote the matrix $\partial_a\partial_b\cG$ as
$ \Sigma_{ab} \equiv G_{ab} + i B_{ab}$.
The holomorphic 3-form $\Omega$ can be written by this basis as
\bea
\Omega = z^a \omega_a.
\ena
Since any vector of $H^{(3,0)} \oplus H^{(2,1)}$ can be written in the form
$C = C^a \omega_a$, a natural inner product of these basis can be defined as
\bea
<C \vert D> = {i\over 2} \int_\cM C \w \bar{D} = C^a B_{ab}\bar{D}^b,
\ena
In other words, a natural metric of the vectors are imaginary part of the
period matrix $ \Sigma_{ab}$.  These inner products are the basic set-up for
the special geometry whereas the basis $(\a_a,\b_a)$ can be called integral
basis\cite{Special}.

The most important ingredient of our discussion is Hodge $*$ operator. Hodge
$*$ operation maps the 3-forms to 3-forms. We write the effect on the basis as
\bea
* \a_a &=& - f_a^b \a_b + e_{ab}\b^b,\nonumber\\ {}*\b^a &=& - g^{ab}\a_b +
f_b^a  \b^b,
\ena
so that we have
\bea
\int_\cM \a_a \w *\a_b &=& e_{ab}, \quad \int_\cM \b^a \w *\b^b = g^{ab},
\nonumber\\
\int_\cM \a_a \w *\b^b &=& \int \b^b \w *\a_a = f_a^b.
\ena
Since the relation $**\gamma = - \gamma $ should hold for any 3-form $\gamma$,
these coefficients satisfy the following relations;
\bea
e_{ac}g^{cb} - f_a^c f_c^b &=& \delta_a^b,\nonumber\\
 f_a^c e_{ab} = f_b^c e_{ac}&,& \quad f_c^b g^{ac} = f_c^a g^{cb}.
\ena

Let us determine the coefficients $e_{ab},f_a^b,g^{ab}$.
These coefficients can be completely determined by $z^a$. Note that
the effect of $*$-operator on $(3,0)$ form $\Omega$ is
\bea
* \Omega = -i \bar{\Omega},
\ena
whereas on $(2,1)$ form $C$ the operation acts as
\bea
* C = i \bar{C}.
\ena
This difference is suffice to determine the coefficients.
The basis of $(2,1)$ form should be perpendicular to $\Omega$ with respect to
the inner product defined in $(22)$.  These are written as
\bea
\omega_a - {B_{ab}\bar{z}^b \over <z \vert z>}\Omega,
\ena
where $<z \vert z> = B_{ab}z^a\bar{z}^b$ as defined in $(22)$.
We require that $*$-operator acts on the function of the moduli as complex
conjugation to get real number for the norm\footnote{We should say that
another expression of the operation can be obtained without this requirement. I
am not convinced that this this is the proper choice  at present}.
By acting  the star operator on $(23)$, and by separation the real and
imaginary part, we obtain
\bea
e_{ab} &=&  { 1 \over <z \vert z>}  (\cG_a \bar{\cG}_b + \bar{\cG}_a \cG_b)-
G_{ac}(B^{-1})^{cd} G_{db} - B_{ab} , \nonumber\\
f_{a}^b &=&  { 1 \over <z \vert z>}(\cG_a \bar{z}^b + \bar{\cG}_a z^b)-
G_{ac}(B^{-1})^{cb} , \nonumber\\
g^{ab} &=&  { z^a \bar{z}^b + \bar{z}^a z^b \over <z \vert z>}-(B^{-1})^{ab} .
\ena

After preparing the mathematical formula, we are now going to derive the mass
formula of the black hole coming from the compactification of the self-dual
5-form field strength.  We expand the 5-form $F$ by
\bea
F = F^a \a_a + G_a \b^a,
\ena
where $F^a$ and $G_a$ are 2-form field strength with respect to our space-time.
Then the action can be written as
\bea
\int_{\cM \times M^4} F * F =  \int_{M^4}[e_{ab}F^a \w * F^b + 2 f_a^bF^a\w
*G_b + g^{ab}G_a \w *G_b].
\ena
We can therefore regard the parameters $e_{ab}, f_a^b$ and $g^{ab}$ as coupling
constants. These coupling constants are functions of the muduli parameters as
was shown  in $(28)$.
Equations of motions are satisfied when $dF^a = dG_a = d*F^a = d*G_a = 0$.
We consider Reisner-Nordstrom solution and impose  the quantization conditions
on the magnetic component as
\bea
\int_{A^a \times S^2} F &=& C_g n^a, \nonumber\\
\int_{B_a \times S^2} F &=& C_g m_a.
\ena
The corresponding fields may be regarded as the ones having an elemental charge
$C_g$ and the  winding number $(n^a,m_b)$ with respect to the internal
manifold.
The magnetic component can be solved by taking
$(F_m)^a = C_g n^a \sin\theta d\theta \w d \varphi, \quad (G_m)_a = -C_g m_a
\sin\theta d \theta \w d \varphi $
Then the solution of the 5-form $F$ can be given by
\bea
F = (F_m)^a \w \a_a + (G_m)_a \w \b^a + *(F_m)^a \w *\a_a + *(G_m)_a \w *\b^a.
\ena
We can now derive the mass formula of the extreme black hole. The mass is the
sum of the contributions of the electric charges and those of magnetic charges.
These contribution should be identical by electromagnetic duality..
As a matter of fact, When we denote the electric charge and the magnetic charge
of $F^a$ and $G_b$ by $Q^a,P^a$ and $Q_b,P^b$ respectively, the solution
$(31)$implies the electric charge can be written as
\bea
P^a = f_b^aQ^b - g^{ab}Q_b,\quad P_a = e_{ab}Q^b + f_a^b Q_b.
\ena
Using this relation, we find the following identity;
\bea
e_{ab}Q^aQ^b &+& 2f_a^bQ^aQ_b + g^{ab}Q_aQ_b \nonumber\\
&=& e_{ab}P^aP^b + 2f_a^bP^aP_b + g^{ab}P_aP_b.
\ena
The final  formula of the extreme black hole is given by
\bea
M^2 = {2C_g^2 \over Vol{\cM}}[e_{ab}n^an^b -2f_a^bn^am^b + g^{ab}m_am_b],
\ena
where $Vol{\cM}$ is the volume of the Calabi-Yau space coming from the fact
that the Einstein action aquires this factor.

Note that this formula is different from an generalization of BPS mass
formula\cite{CDFV}
\bea
M = v \vert z^am_a - \cG_a n^a \vert
\ena
where
corresponding $e_{ab},f_a^b$ and $g^{ab}$ are identified as
\bea
e_{ab} &=& \cG_a \bar{\cG}_b + \bar{\cG}_a \cG_b,\nonumber\\
f_a^b &=& (z^a\bar{\cG}_b + \bar{z}^a \cG_b), \nonumber\\
g^{ab}&=& z^a\bar{z}^b + \bar{z}^a z^b.
\ena

At present, the derived formula $(36)$ is merely classical. When we proceed
further, we should consider the quantum behavior of the coupling constants. It
might be that the classical result contain the information of the loop
correction.\cite{Strominger,GMS} It seems  also instructive to consider models
where  the effective coupling constants can be calculated explicitly.


\newpage
\begin{thebibliography}{99}

\bibitem{Strominger}
A. Strominger,
" Massless Black Holes and Conifolds in String Theory," hep-th 9504090.


\bibitem{SeibergWitten}
N. Seiberg and E. Witten,
Nucl. Phys. B426 (1994) 19.

\bibitem{GMS}
B. R. Greene, D. R. Morrison and A. Strominger,
"Black Hole Condensation and the Unification of String Vacua," CLNS-95/1335,
hep-th 9504145.

\bibitem{HS}
G. Horowits and A. Strominger,
Nucl. Phys. B360 (1991) 197.

\bibitem{GH}
G. W. Gibbons and C. M. Hull,
Phys. Lett. 109B (1982) 190.


\bibitem{CDFV}
A. Ceresole, R. D'Auria, S. Ferrara and A. Van Proeyen,
"Duality Transformations in Supersymmetric Yang-Mills Theory Coupled to
Supergravity," hep-th 9502072.

\bibitem{CGH}
P. Candelas, P.S. Green and H. H{\"u}bsch,
Nucl. Phys. B330 (1990) 49.

\bibitem{BG}
R. Bryant and P. Griffiths,
Progress in Mathematics 36, pp.77, (Birk{\"o}user, Boston, 1983)

\bibitem{Special}
A. Strominger, Comm. Math. Phys. 133 (1990) 163.

\end{thebibliography}
\end{document}